\documentclass[superscriptaddress,twocolumn,aps,pra,showpacs,preprintnumbers,amsmath,amssymb]{revtex4}
\usepackage{epsfig}
\usepackage{amssymb,amsfonts}
\usepackage{bm}
\usepackage{ulem}
\usepackage{epsfig}
\usepackage{graphicx}
\usepackage{amssymb,amsmath,amsbsy,amsgen,amsfonts}
\usepackage{dcolumn}
\usepackage{amsthm}
\usepackage{mathrsfs}
\usepackage{latexsym}
\usepackage{array}
\usepackage{color}
\usepackage{amstext}
\allowdisplaybreaks[1]
\usepackage{txfonts}

\usepackage{epstopdf} %for texshop on mac

\newcommand{\bra}[1]{\langle{#1}\vert}
\newcommand{\ket}[1]{\vert{#1}\rangle}

\newcommand{\be}{\begin{equation}}
\newcommand{\ee}{\end{equation}}
\newcommand{\ba}{\begin{array}}
\newcommand{\ea}{\end{array}} 
\newcommand{\bqa}{\begin{eqnarray}}
\newcommand{\eqa}{\end{eqnarray}}

\DeclareSymbolFont{symbols}{OMS}{cmsy}{m}{n}

\begin{document}
\def\mydag{^{\vphantom{\dagger}}}
%\title{Optical emulation of dissipation-induced correlations in waveguide arrays}
\title{Photonic lattice simulation of dissipation-induced correlations in bosonic systems}

\author{Amit Rai}
\affiliation{Centre for Quantum Technologies, National University of Singapore, 3 Science Drive 2, Singapore 117543}
\affiliation{Department of Physics and Astronomy, National Institute of Technology, Rourkela, 769008 India}
\author{Changhyoup Lee}
\email{changdolli@gmail.com}
\affiliation{Centre for Quantum Technologies, National University of Singapore, 3 Science Drive 2, Singapore 117543}
\author{Changsuk Noh}
\email{cqtncs@nus.edu.sg}
\affiliation{Centre for Quantum Technologies, National University of Singapore, 3 Science Drive 2, Singapore 117543}
\author{Dimitris G. Angelakis}
\email{dimitris.angelakis@gmail.com}
\affiliation{Centre for Quantum Technologies, National University of Singapore, 3 Science Drive 2, Singapore 117543}
\affiliation{School of Electronic and Computer Engineering, Technical University of Crete, Chania, Greece 73100}

\date{\today}

\begin{abstract}
We propose an optical simulation of dissipation-induced correlations in one-dimensional (1D) interacting bosonic systems, using a two-dimensional (2D) array of linear photonic waveguides and only classical light. %\textcolor{blue}{In the presence of} two-body losses  and
%Considering two-body losses, 
We show that for the case of two bosons in a 1D lattice, one can simulate on-site two-body dissipative dynamics using a linear 2D waveguide array with lossy diagonal waveguides. 
%The proposed photonic set-up allows direct visualisation of the wave function, enabling one to observe the dissipation-induced correlations. 
The intensity distribution of the propagating light directly maps out the wave function, allowing one to observe the dissipation-induced correlations with simple measurements.
Beyond the on-site model, we also show that a generalised model containing nearest-neighbour dissipative interaction can be engineered and probed in the proposed set-up.
%The latter model shows the versatility of photonic lattices emulator.
\end{abstract}


\maketitle
{\it Introduction.---} 
Photonic lattices, an array of optical waveguides, have recently emerged as a successful experimental platform to emulate diverse physical phenomena. 
Most of the works in this field focus on single-particle phenomena, where examples include optical Bloch oscillations of various kinds\cite{Peschel,Pertsch,MorandottiSilberberg1999,Longhi12a,Longhi12c}, continuous-time random walks\cite{quantumwalk1}, Anderson localisation\cite{anderson1,anderson2,Martin}, dynamic localisation\cite{Szameit10}, and dynamic band collapse\cite{Crespi13}. Simulations of relativistic equations and related effects\cite{Longhi11c}, such as photonic \textit{Zitterbewegung}~
\cite{Dreisow}, Klein tunneling\cite{Dreisow2012}, and random mass Dirac model\cite{Keil13} have also been performed, including the simulation of unphysical Majorana equation\cite{Keil14}.
However, it has been shown that phenomena involving more than one particle can also be simulated in waveguide arrays\cite{Longhi10a,Longhi11a}. In particular, one can simulate the physics of two interacting particles using two-dimensional (2D) square arrays of linear waveguides along with classical light\cite{LonghiValle2011,KrimerKhomeriki2011}, allowing even richer physics such as Bloch oscillations of correlated particles\cite{Longhi11b}, fractional Bloch oscillations\cite{fractionalblochoscillation}, and Anderson localisation of two interacting bosons\cite{CLEE} to be observed in photonic lattices.

In all of the examples above, dissipation is either an adverse effect that destroys the relevant effect or one that does not play a significant role. However, recent studies have shown that decoherence or dissipation can actually be the main source of non-trivial quantum effects.
In the case of optical systems, losses have been deliberately introduced to realise parity-time symmetric systems\cite{El-Ganainy,Longhi09,Guo,Ruter}, whereas in an optomechanical system it was shown that it is possible to generate the squeezed state by using dissipation\cite{Kronwald}. In optical lattices, strong inelastic collisions were used to inhibit particle losses and drive the system into a strongly correlated regime\cite{Syassen, Durr, Garcia}.  
There, the two-body inelastic collisions are induced by creating molecules using Feshbach resonance, whereas one-body losses are negligible due to the stability of the system and the absence of thermal background of particles.
%\textcolor{red}{In the latter, the inelastic collisions result in two-particle losses, which act as an effective repulsion between particles and thereby reduce the overall loss and induce correlations.}
%

In this work, we show for the first time that an essential part of such dissipation-induced physics can be simulated using a linear 2D waveguide structure, and moreover using only classical light. Our proposed waveguide simulator allows for highly-tunable effective two-body dissipation rate while having no effective single body losses, making it an excellent candidate to simulate the non-trivial physics induced by strong two-body dissipation.
We first introduce a connection between the photonic lattice system and two-body dissipative Bose-Hubbard system, which holds in the two-particle sector.  We then discuss how the proposed system allows visualisation of the wave function and relevant observables, and use the fact to illustrate dissipation-induced physics. Interestingly, we find that an effective Hamiltonian description is completely equivalent to the master equation description in the proposed system. The versatility of the proposed set-up is highlighted by introducing a generalised model that goes beyond the on-site dissipative Bose-Hubbard model, whose signatures are briefly examined. 

%\section{Results}

\begin{figure}
\centering
\includegraphics[width=8cm]{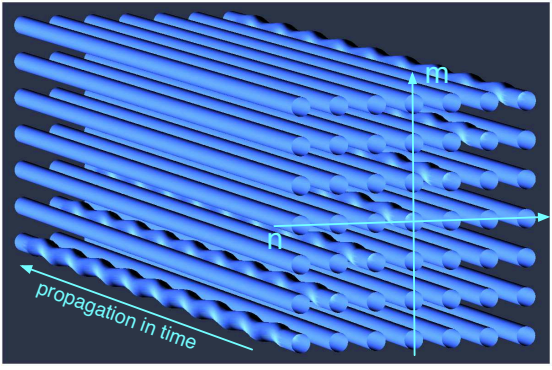}
\caption{\label{waveguide}
Schematic diagram of the proposed 2D waveguide array to simulate the 1D lattice system with on-site two-body losses, where sinusoidal modulations at the diagonal waveguides are introduced to induce controllable radiation losses, while the off-diagonal waveguides are assumed to be uniform, i.e., lossless.}
\end{figure}

%\subsection{Proposed photonic lattice system.}
{\it Proposed photonic lattice system.---} 
Our proposal relies on a mapping between a 2D square waveguide array (Fig.~\ref{waveguide}) and one-dimensional (1D) Bose Hubbard model (BH) in the two-particle sector\cite{LonghiValle2011,KrimerKhomeriki2011,fractionalblochoscillation}. 
The light propagation in a symmetric square 2D waveguide array can be described by the coupled-mode equations:
\begin{eqnarray}\label{twod}
\begin{split}
i \dot{c}_{n,m}
 = \beta \hspace{0.05cm} \delta_{n,m} {c}_{n,m} -{\kappa}(\hspace{0.05cm}
{c}_{n,m+1} + {c}_{n,m-1}+{c}_{n+1,m}+{c}_{n-1,m})
\end{split}
\end{eqnarray}

\noindent where $c_{n,m}(t)$ describe the amplitudes of the classical field at site $(n,m)$; $\beta$ is a shift in the propagation constant of the diagonal waveguides compared to that of the off-diagonal waveguides; $\kappa$ is the evanescent coupling strength between neighbouring waveguides. The propagation constants of the off-diagonal waveguides are set to zero for convenience; 
small intrinsic waveguide losses will be ignored in this work, as they merely rescale the total intensity along the propagation distance.
In general, $\beta$ takes a complex value
\begin{equation}
\beta = \beta_{r}-i~\Gamma,
\end{equation}
where $\beta_r$ is the phase constant and $\Gamma$ is the attenuation constant. Normally, losses are neglected in photonic lattice systems, but recently it was shown that losses can be controllably induced by modulating the waveguides in the transverse direction\cite{Eichelkraut} in a 1D waveguide array, where the loss rate of up to $\Gamma/\kappa=10$ has been demonstrated. In our proposed set-up, $\Gamma$ is introduced by transverse modulation of the diagonal waveguides in the horizontal/vertical plane as shown in Fig.~\ref{waveguide}.

To see the connection between this system and the BH system, consider the following non-Hermitian Hamiltonian of the BH type:
\begin{equation}
\hat{H}_\textrm{NHBH}=  - \kappa \sum_{j}  (\hat{a}_{j}^{\dagger} \hat{a}_{j+1}+\hat{a}_{j+1}^{\dagger} \hat{a}_{j}) +  \frac{\beta}{2}\sum_{j}\hat{n}_{j}(\hat{n}_{j}-1).
\label{BHH}
\end{equation}
Writing the state as,
\begin{align}
\ket{\psi (t)} = \frac{1}{\sqrt{2}}\sum _{n,m} c_{n,m}(t)\hat{a}_{n}^\dagger \hat{a}_{m}^\dagger \ket{0},
\label{mapping}
\end{align}
the corresponding Schr\"odinger equation reduces to Eq.~(\ref{twod})\cite{fractionalblochoscillation}. 
%\textcolor{red}{Note that  $c_{n,n}$ ($\sqrt{2}c_{n,m}$) is the probability amplitude of finding the bosons at site $n$ (at sites $n$ and $m$), where $c_{n,m} = c_{m,n}$ is enforced.}
Thus, the light amplitude in the diagonal (off-diagonal) waveguide, $c_{n,n}$ ($\sqrt{2}c_{n,m}$), corresponds to the probability amplitude of finding the two bosons at the same site $n$ (at different sites $n$ and $m$). 
For $\Gamma=0$, the 2D waveguide array therefore becomes a photonic emulator of the two-particle BH model with hopping rate $\kappa$ and on-site nonlinearity $\beta_r$. 
For $\Gamma\neq 0$, the model contains an effective two-body loss term that mimics the inelastic two-body collision of cold atoms. 
Also note that because of Eq.~(\ref{mapping}), the simulated Hilbert space of the proposed waveguide set-up always stays in the two-body manifold of the non-Hermitian BH model. Thus, by construction, unwanted one-body losses are absent in our proposal. Light losses in an off-diagonal waveguide, if induced, correspond to a long-range two-body dissipative interaction. We will utilise this fact later to generalise the on-site interaction model to the nearest-neighbour interaction model.

%\textcolor{red}{Here we would like to point out that the above mapping enables the 2D waveguide array to map out the two-boson manifold, i.e., when there are exactly two bosons in the original 1D lattice system. 
%This implies that one-body losses are naturally ignored in the proposed waveguide array simulator since the mapping does not contain one-body manifold. 
%Losses in the off-diagonal waveguides correspond to long-range dissipation interactions between bosons and are considered later when we generalise the non-Hermitian BH model (\ref{BHH}).
%}

A proper description of the simulated lossy BH system in Eq.~(\ref{BHH}) requires the master equation formalism, where the above effective Hamiltonian description is only valid for a short time evolution. However, as we show later, the effective Hamiltonian is exactly equivalent to the master equation description in our two-particle problem.

%%%%%%%%%%%%%%%%%%%%%%%%%%%%%%%%%%%%%%%%%%%%%%%%%%%%%%%%%%%%%%%%%%%%%%%%
%%%%%%%%%%%%%%%%%%%%%%%%%%%%%%%%%%%%%%%%%%%%%%%%%%%%%%%%%%%%%%%%%%%%%%%%
%\subsection{Visualisation of the wavefunction and observables.}
{\it Visualisation of the wavefunction and observables.---} 
One of the most attractive features of photonic lattice simulators is their ability to visualise a wave function under study. Equation (\ref{mapping}) provides a direct link between the classical field amplitudes of the waveguide array and the wave function of two bosons, which also enables preparation of an arbitrary initial state with classical sources\cite{CLEE}.
Here, the measured intensities $\vert c_{n,n}\vert^2$ ($2\vert c_{m,n}\vert^2$) correspond to the probabilities to find the bosons at site $n$ (at sites $m$ and $n$).

In this work, we use the average particle number and average intensity correlations to describe dissipation-induced inhibition of losses and correlations. These quantities only require intensity distributions and therefore are experimentally accessible.
The average particle number remaining is defined as $N_\textrm{tot}(t) = \sum_k N_k(t)$, where $N_k(t)$ is the normalised particle-density distribution
\begin{align}
N_k (t) \equiv \frac{1}{2}\bra{\psi(t)}\hat{n}_{k}\ket{\psi(t)}=\sum_{n}\vert c_{k,n}(t)\vert^{2},
\end{align}
with $\hat{n}_k$ the particle number operator at site $k$. 
The average intensity correlations $G^{(2)}_\textrm{avg}(t) \equiv \sum_{n} G^{(2)}_{n,n}(t)/L$, and its normalised version $g^{(2)}_\textrm{avg}(t) \equiv \sum_{n} g^{(2)}_{n,n}(t)/L$, are defined via 
\begin{eqnarray}
G^{(2)}_{n,m}(t) = \langle \hat{a}^\dagger_n \hat{a}^\dagger_m\hat{a}_m\hat{a}_n\rangle = 2 \vert c_{n,m}(t)\vert^2,
\label{G2nm}
\end{eqnarray}
and
\begin{eqnarray}
g^{(2)}_{n,m}(t) = \frac{G^{(2)}_{n,m}(t)}{\langle \hat{n}_n\rangle\langle\hat{n}_m\rangle},
\end{eqnarray}
where $L$ is the total number of sites in the 1D lattice.

%%%%%%%%%%%%%%%%%%%%%%%%%%%%%%%%%%%%%%%%%%%%%%%%%%%%%%%%%%%%%%%%%%%%%%%%
%%%%%%%%%%%%%%%%%%%%%%%%%%%%%%%%%%%%%%%%%%%%%%%%%%%%%%%%%%%%%%%%%%%%%%%%
%\subsection{Effective Hamiltonian description.}
{\it Effective Hamiltonian description.---} 
Here, we first provide the proper master equation description for the dissipative (non-Hermitian) BH system introduced above, which holds for any number of particles. We then prove an equivalence between the master equation and the Schr\"odinger equation with non-Hermitian Hamiltonian (\ref{BHH}) for the two-particle case.

In the presence of losses, the quantum state no longer stays pure and must be described by a density operator. The Liouville equation can be written in the Lindblad form, which for the two-body loss case yields\cite{Garcia}
\begin{equation}
\dot{\rho} =  - \frac{i}{\hbar} [H_\textrm{BH}, \rho] +\frac{\Gamma}{2} \sum_{j} (2 \hat{a}_{j}^2 \rho \hat{a}_{j}^{\dagger 2} - \hat{a}_{j}^{\dagger2}  \hat{a}_{j}^2 \rho- \rho \hat{a}_{j}^{\dagger2}  \hat{a}_{j}^2),
%\label{mastereqn}
\end{equation}
with the usual BH Hamiltonian $H_\textrm{BH}$ (the lossless version of Eq.~(\ref{BHH}), i.e., $\beta=\beta_r$). 
In the short time limit, the quantum `jump' term,  $\hat{a}_{j}^2 \rho \hat{a}_{j}^{\dagger 2}$, can normally be ignored\cite{Durr,Carmichael}, in which case the master equation reduces to an effective Schr\"odinger equation with the non-Hermitian Hamiltonian (\ref{BHH}):
\begin{equation}
\frac{d\rho}{dt} = - \frac{i}{\hbar} \left[ \hat{H}_{\rm NHBH}, \rho \right].
\end{equation}
However, in the case of two particles, an analysis using the master equation is already equivalent to that using an effective Schr\"odinger equation, since there is no channel into the single particle manifold but only an incoherent channel to the vacuum. The latter changes the overall probability to be in the two-particle manifold, but not the two-particle state itself. Therefore $\rho(t) = P_2(t) \ket{\psi_{\rm TP}(t)} \bra{\psi_{\rm TP}(t)} + P_0(t) \ket{0} \bra{0}$, where $\ket{\psi_{\rm TP}}$ is the state in the two-particle manifold whose dynamics is governed by $\hat{H}_{\rm NHBH}$ and $P_{2}(t)$ $(P_{0}(t))$ is a probability to be in the two-particle (vacuum) manifold. Therefore, if one is only interested in physics captured by the two-particle sector, the master equation is exactly equivalent to the non-Hermitian evolution under the Schr\"odinger equation.

%%%%%%%%%%%%%%%%%%%%%%%%%%%%%%%%%%%%%%%%%%%%%%%%%%%%%%%%%%%%%%%%%%%%%%%%
%%%%%%%%%%%%%%%%%%%%%%%%%%%%%%%%%%%%%%%%%%%%%%%%%%%%%%%%%%%%%%%%%%%%%%%%
%\section{Observation of dissipation-induced physics}

\begin{figure}[b]
\centering
\includegraphics[width=8.7cm]{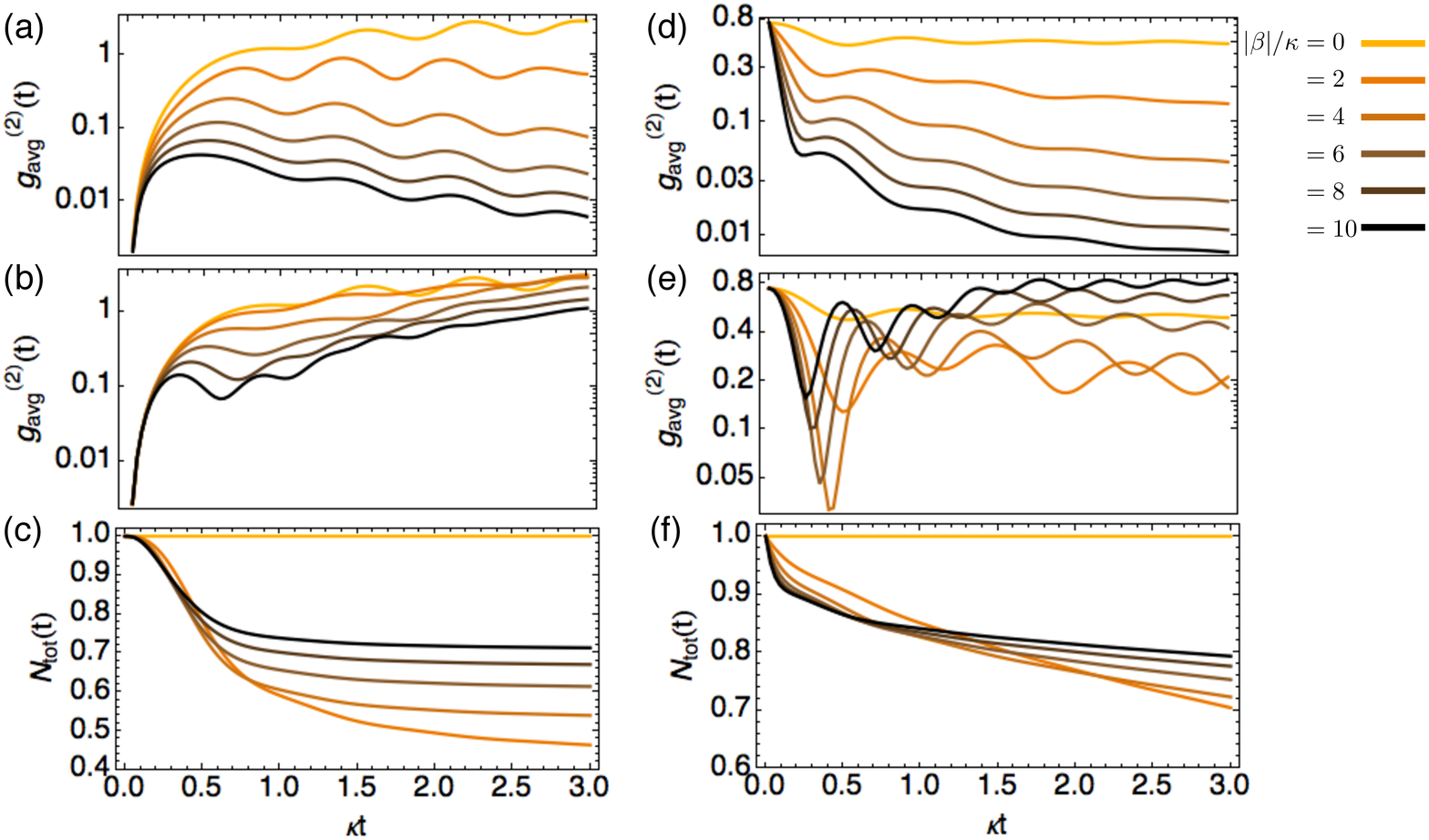}
\caption{\label{dynamics}
Intensity correlation $g^{(2)}_\textrm{avg}(t)$ as a function of time under dissipative (top row) and unitary dynamics (middle row) for the localised (left column) and homogeneous (right column) initial state. 
Bottom row: normalised total number of bosons as a function of time. 
Note that stronger dissipation yields stronger final photon antibunching. For the same values of $\vert \beta \vert/\kappa$, the unitary nonlinear interaction strength $\beta_{r}/\kappa$ 
exhibits different behaviours.
%tends to \textcolor{red}{induce photon bunching}. \textcolor{blue}{(Can we say `bunching'?) \textcolor{green}{I think so, because it eventually goes over 1.}} 
For both types of initial states, stronger dissipation results in lower overall loss.}
\end{figure}

%\subsection{Dissipation-induced strong correlations and inhibition of loss.}
{\it Dissipation-induced strong correlations and inhibition of loss.---} 
Unlike the usual single-particle dissipation, two-body dissipation by itself can give rise to interesting physical effects. For example, particles tend to stay away from each other to reduce dissipation and in the process create strong correlations\cite{Syassen}. Here, we show that these effects can be simulated and observed in the proposed set-up. For this purpose, we first consider a localised initial state $\ket{\psi(0)} = \hat{a}_{0}^\dagger\hat{a}_{1}^\dagger\ket{0}$ propagating in the 15-site lattice (15 by 15 2D waveguide lattice, i.e., $L=15$). The average same-site intensity correlation function $g^{(2)}_\textrm{avg}(t)$ for the cases of purely dissipative and purely unitary dynamics are shown in Fig.~\ref{dynamics} (a) and (b). Upon comparison, the effects of the dissipative dynamics is clear. In the unitary case, the correlation function builds up continuously with time, whereas in the dissipative case, it ultimately decreases with increasing dissipation rate. Note that while the unitary interaction does keep the correlation at bay, the effect is much weaker than the dissipative interaction. We have checked that the required strength of the unitary interactions to achieve similar final correlations to the dissipative case is 10 times larger. Importantly, the induced (anti-)correlations are accompanied by inhibition of losses as shown in Fig.~\ref{dynamics} (c), signifying that the observed (anti-)correlations did not arise from the fact that particles have dissipated away. In fact, the remaining fraction increases with dissipation rate, for instance, from $45\%$ to $70\%$ for from $\Gamma/\kappa=2$ to $\Gamma/\kappa=10$.

In the above example, we have used an initial state with $g^{(2)}_\textrm{avg}(0)=0$ and a localised (inhomogeneous) distribution. To study the dissipation-induced effects on a homogeneous initial state that has nonzero $g^{(2)}_\textrm{avg}(0)$, we consider the superposition of a homogeneous two-site occupied state $\ket{TS} = \sqrt{\frac{2}{L(L-1)}} \sum_{i < j} \hat{a}_i^\dagger\hat{a}_j^\dagger \ket{0}$ and a single-site occupied state $\ket{SS} = \frac{1}{\sqrt{2L}}\sum_i \hat{a}_i^\dagger\hat{a}_i^\dagger \ket{0}$ with weights $\alpha_{TS}$ and $1-\alpha_{TS}$ respectively, i.e., $\ket{\psi(0)} = \sqrt{\alpha_{TS}}\ket{TS} + \sqrt{1-\alpha_{TS}}\ket{SS}$.
As an example, results for the case of $\alpha_{TS}=9/10$ are displayed in Fig.~\ref{dynamics} (d), (e), and (f) for the same values of $\beta$ used in the case of the local initial state. This nonlocal initial state has non-zero $g^{(2)}_\textrm{avg}(0) = (1-\alpha_{TS})L/2$, leading to a rapid initial dissipation followed by a slower decay that decreases with increasing $\Gamma/\kappa$. The established final correlation (antibunching) increases with the dissipation rate, while it generally decreases with the unitary interaction strength $\beta_r$. Because the initial correlation function takes a  non-zero value, the (anti-)correlation can be said to have been induced by the dissipative dynamics.

\begin{figure}[t]
\centering
\includegraphics[width=8.5cm]{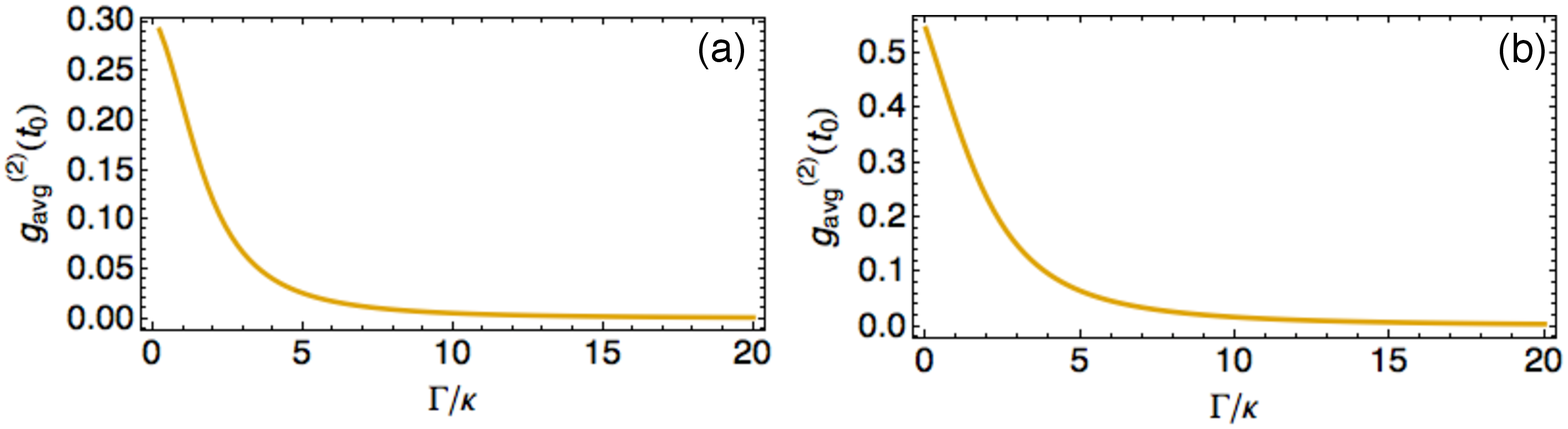}
\caption{\label{g2vsU}
Intensity correlation function $g^{(2)}_\textrm{avg}(t_{0})$ at time $\kappa t_0 = 1$ as a function of $\Gamma/\kappa$ for a) the localised state and b) the homogeneous state.}
\end{figure}

The behaviour of the correlation function $g^{(2)}_\textrm{avg}(t_{0})$ at a fixed time $t_{0}$ as a function of $\Gamma/\kappa$ is similar for both the local and homogeneous initial states as shown in Fig.~\ref{g2vsU}. It decreases rapidly with $\Gamma/\kappa$ and becomes almost 0 for $\Gamma/\kappa> 10$, i.e., the larger the loss-to-coupling ratio, higher the correlations in the final state.

%%%%%%%%%%%%%%%%%%%%%%%%%%%%%%%%%%%%%%%%%%%%%%%%%%%%%%%%%%%%%%%%%%%%%%%%
%%%%%%%%%%%%%%%%%%%%%%%%%%%%%%%%%%%%%%%%%%%%%%%%%%%%%%%%%%%%%%%%%%%%%%%%

\begin{figure}[t]
\centering
\includegraphics[width=8.5cm]{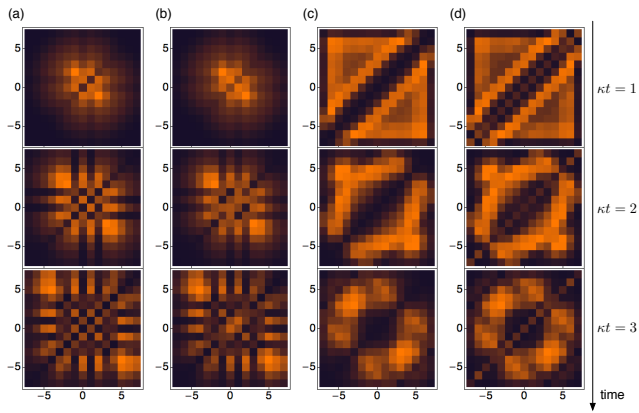}
\caption{\label{visualizationBH}
Time-evolution of the cross-correlation function for the local initial state with $\Gamma/\kappa=10$ in (a) and with $\beta_{r}/\kappa=10$ in (b), and for the homogeneous initial state with $\Gamma/\kappa=10$ in (c) and with $\beta_{r}/\kappa=10$ in (d).}
\end{figure}

Cross-correlations can also be observed using the aforementioned ability to visualise the wave function. We thus plot the intensity distribution in the proposed 2D waveguide array in Fig.~\ref{visualizationBH} for the dissipative ((a) and (c)) and the unitary interaction cases ((b) and (d)), respectively. 
The two left columns are for the localised initial state whereas the two right columns are for the homogeneous initial state. The absence of diagonal elements in the dissipative case displays the tendency for bosons to stay apart from each other. On the contrary, the diagonal waveguides are clearly occupied for the unitary interaction case, giving rise to the significant average correlation function as shown earlier. The off-diagonal elements exhibit very similar distributions due to the local nature of the interaction, although there is a slight enhancement near anti-diagonal elements for the dissipative case. Due to the nature of the mapping, the intensity distribution directly images the unnormalised cross-correlation function $G^{(2)}_{n,m}$, providing a good experimental probe of the dissipative-induced correlations. 

%%%%%%%%%%%%%%%%%%%%%%%%%%%%%%%%%%%%%%%%%%%%%%%%%%%%%%%%%%%%%%%%%%%%%%%%
%%%%%%%%%%%%%%%%%%%%%%%%%%%%%%%%%%%%%%%%%%%%%%%%%%%%%%%%%%%%%%%%%%%%%%%%

%\subsection{Generalisation to extended model with nearest-neighbour dissipative interaction}
%\subsection{Beyond on-site dissipation.}
{\it Beyond on-site dissipation.---} 
\begin{figure}[t]
\centering
\includegraphics[width=5cm]{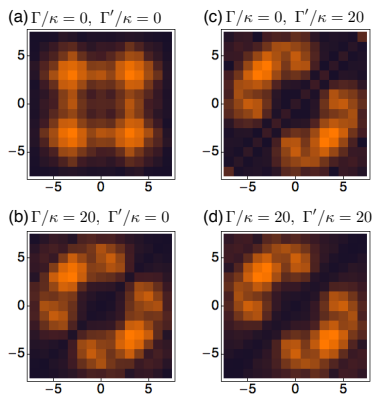}
\caption{\label{visualizationEBH}
Normalised cross correlation function of the initial homogeneous state at $\kappa t_0 = 3$ for various values of $\Gamma/\kappa$ and $\Gamma'/\kappa$. Depending on the type of dissipation, the tri-diagonal waveguides exhibit various behaviours.}
\end{figure}
The 2D waveguide array allows one to go beyond the dissipative BH model and simulate an extended dissipative BH model, where the nearest-neighbour (NN) dissipation is included:
\begin{align}
\hat{H}_{\rm ext} = \hat{H}_{\rm NHBH}  -i~\Gamma' \sum_j \hat{n}_j\hat{n}_{j+1}.
\end{align}
The NN dissipation $\Gamma'$ can be realised by the modulation of the NN ($m=n\pm1$) diagonal waveguides in the proposed 2D waveguide array.
This type of long-range dissipative interaction is usually absent in bosonic systems and the ability to implement such a term demonstrates the strength of the proposed waveguide system. The extra interaction term brings with it richer physics, a part of which is discussed briefly here.
%To our knowledge, this is the first proposal to realise such long-range dissipative interaction, which is allowed due to the versatility of the photonic lattice system.
%Such long-range dissipative interaction is usually readily in other systems.
%Figure \ref{visualizationEBH} depicts the effects of the additional NN dissipation term. 

Figure \ref{visualizationEBH} depicts the cross correlations developed in the time evolution under the extended dissipative BH model.
The left column shows the cross correlation functions of the previously studied non-dissipative and the on-site dissipative cases for initially homogeneous state. Figure \ref{visualizationEBH}(c) shows the NN-dissipative case, where only the correlation function between the NN sites are suppressed, visualised by vanishing intensities in the waveguides directly above and below the diagonal waveguides. 
Finally, when $\Gamma$ and $\Gamma'$ are both non-zero, both the on-site and NN correlations are suppressed.  The latter two are new types of correlated bosonic states created by the unique extended dissipative BH model whose simulation is allowed naturally by the proposed waveguide array set-up.

%%%%%%%%%%%%%%%%%%%%%%%%%%%%%%%%%%%%%%%%%%%%%%%%%%%%%%%%%%%%%%%%%%%%%%%%
%%%%%%%%%%%%%%%%%%%%%%%%%%%%%%%%%%%%%%%%%%%%%%%%%%%%%%%%%%%%%%%%%%%%%%%%

%\section{Discussion}
{\it Discussion.---} 
In conclusion, we have shown that it is possible to use classical light propagation in two dimensional arrays of optical waveguides to simulate dissipation-induced strong correlation effects. The proposed photonic lattice system has lossy waveguides along the diagonal, whose loss rates can be controlled by introducing transverse modulation in the diagonal axis. We proved that the two-body lossy system can be described by an effective Hamiltonian, instead of a master equation, for any two-particle initial states. This implies that the 2D photonic lattice system is a faithful simulator of the investigated system. We showed that observables such as the intensity correlation functions and normalised particle density distribution can be measured experimentally, providing direct probes of the simulated dissipation-induced phenomena. In particular, the ability to visualise the wave function helps in observing the induced correlations. Lastly, we have proposed and studied an extended dissipative BH model where nearest-neighbour dissipative interaction is included. Further investigations into this model and towards its realisation in other platforms provide an interesting avenue for future research.

%\section{Acknowledgments.}
{\it Acknowledgments.---} 
We would like to acknowledge the financial support provided by the National Research Foundation and Ministry of Education Singapore (partly through the Tier 3 Grant ``Random numbers from quantum processes"), and travel support by the EU IP-SIQS.

\end{document}